\documentclass[12pt]{iopart}
\usepackage[dvips,final]{graphicx}
\usepackage{latexsym}
\begin{document}

\title[The Accelerating Universe and a Limiting Curvature Proposal ]{The Accelerating Universe and a Limiting Curvature Proposal }

\author{Damien A. Easson}

\address{Centre for Particle Theory, Department of Mathematical Sciences, Durham University,
Science Laboratories, South Road, Durham, DH1 3LE, U.~K.}
\ead{damien.easson@durham.ac.uk}
\begin{abstract}
We consider the hypothesis of a limiting minimal curvature in gravity as a way to construct a
class of theories exhibiting late-time cosmic acceleration. Guided by the minimal curvature conjecture (MCC) we 
are naturally lead to a set of scalar tensor theories in which the scalar is non-minimally coupled both to gravity and to the matter Lagrangian. 
The model is compared to the Lambda Cold Dark Matter concordance model and to the observational data using the
 ``gold" SNeIa sample of Riess et.~al.~(2004). An excellent fit to the data is achieved. We present  a toy model designed to demonstrate that such
a new, possibly fundamental, principle may be responsible for the recent period of cosmological acceleration. Observational
constraints remain to be imposed on these models.
\end{abstract}

\pacs{98.80.-k, 04.50.+h, 95.36.+x \\
arXiv: \tt astro-ph/0608034 \rm}
\vspace{2pc}
\noindent{DCPT-06/17}
\maketitle
\def\S{{\mathcal S}}  
\def\I{{\mathcal I}}  
\def\L{{\mathcal L}}  
\def\H{{\mathcal H}}  
\def\M{{\mathcal M}}  
\def\N{{\mathcal N}} 
\def\cP{{\mathcal P}} 
\def\R{{\mathcal R}}  
\def\K{{\mathcal K}}  
\def\W{{\mathcal W}} 
\def\mM{{\mathbf M}} 
\def\mP{{\mathbf P}} 
\def\mT{{\mathbf T}} 
\def\mR{{\mathbf R}}
\def\mS{{\mathbf S}}
\def\mX{{\mathbf X}}
\def\mZ{{\mathbf Z}}
\def\half{{1\over2}}  
\def\be{\begin{equation}}
\def\ee{\end{equation}}
\def\bea{\begin{eqnarray}}
\def\eea{\end{eqnarray}}
\def\g{\gamma}
\def\G{\Gamma}
\def\vp{\varphi}
\def\mpl{M_{\rm pl}}
\def\ls{l_{\rm s}}
\def\l{\lambda}
\def\gs{g_{\rm s}}
\def\d{\partial}
\def\sp{\;\;\;,\;\;\;}
\def\spa{\;\;\;}
\def\r{\rho}
\def\dr{\dot r}
\def\dt{\dot\varphi}
\def\e{\epsilon}
\def\m{\mu}
\def\n{\nu}
\def\om{\omega}
\def\tn{\tilde \nu}
\def\p{\phi}
\def\k{\kappa}
\def\vp{\varphi}
\def\r{\rho}
\def\s{\sigma}
\def\t{\tau}
\def\a{\alpha}
\def\b{\beta}
\def\de{\delta}
\def\ls{\ell_{\rm s}}
\def\lmin{\ell_{\rm min}}
\def\lp{\ell_{\rm pl}}
\def\d{\partial}
\def\half{{1\over2}}  
\newcommand{\eq}[1]{equation~(\ref{#1})}
\newcommand{\eqs}[2]{equations~(\ref{#1}) and~(\ref{#2})}
\newcommand{\eqto}[2]{equations~(\ref{#1}) to~(\ref{#2})}
\newcommand{\fig}[1]{Fig.~(\ref{#1})}
\newcommand{\figs}[2]{Figs.~(\ref{#1}) and~(\ref{#2})}
\newcommand{\GeV}{\mbox{GeV}}
 
\section{Introduction}

One of the most profound discoveries of observational physics is that the universe is accelerating in its expansion~\cite{Riess:1998cb,Perlmutter:1998np,Netterfield:2001yq,Halverson:2001yy,Tonry:2003zg,Bennett:2003bz}. There have been many attempts to explain this late-time acceleration, for example, a pure cosmological constant, dark energy associated with some new scalar field and modified gravitational theories, although all current models require some level of fine-tuning and none are considered to be a complete 
explanation~\footnote{For a recent reviews on the subject see~\cite{Nojiri:2006ri,Copeland:2006wr}.}. 
Whatever is responsible for the current acceleration may arise from some completely new physical principle. This is the possibility we
consider in this paper. Our goal is to construct a toy model that represents a late-time accelerating Universe using a new,
possibly fundamental, principle. As our guiding principle, we hypothesize the existence of a {\it minimal curvature \rm} scale in gravity. 

In a Friedmann, Robertson-Walker (FRW) space-time, without cosmological constant $\Lambda$ and
with only standard matter sources such as dust and radiation, the universe will always decelerate as it
expands. One way to avoid this is to add matter to the system that violates the strong energy condition (SEC). In a cosmological context this
violation constitutes the addition of
matter sources satisfying the equation of state $p \le -1/3 \rho$. A second possibility is to explicitly remove flat space-time as a solution to the theory. In this case the vacuum of the theory, which is approached at late times as the energy density in matter fields becomes more
and more dilute, is not Minkowski space-time, but instead an accelerating Universe~\cite{Carroll:2003wy,Carroll:2004de,Easson:2004fq}. To remove flat
spacetime as a solution we hypothesize the existence of a minimal curvature in our underlying fundamental
theory. The simplest example of this is, of course, to introduce a bare cosmological constant into 
General Relativity. However, in principle there may exist many other ways to achieve this result. Indeed, it appears that
many accelerating cosmological models derived from modified gravity theories contain such a minimal curvature~\cite{Nojiri:2003wx,Schuller:2004nn,Schuller:2004rn}.

The idea of a minimal curvature scale in gravity mirrors that of a maximal curvature scale. In the literature many authors have considered
this possibility and used it to remove 
the curvature singularities of General Relativity by bounding curvature invariants from above at the level of the 
classical action~\cite{markov}-\cite{Easson:2003ia}. In the case of singularity removal, it is necessary to bound {\it all} curvature invariants in order to cover all possible physical situations in which such a singularity may occur. 

By contrast, in the case of a minimal curvature approached
at late times in a homogeneous, isotropic universe, symmetry implies that it is only necessary to bound
the Ricci scalar $R$ from below. Hence, unlike in the case of a maximal curvature hypothesis, we shall
see that one may implement a minimal curvature by using a modified Brans-Dicke theory where the Brans-Dicke field
couples non-minimally to the matter Lagrangian.

Within this context we demonstrate that the existence of the minimal curvature (MC) produces a Universe that evolves from a matter dominated period
to an accelerating phase mimicking the $\Lambda$-Cold-Dark-Matter ($\Lambda$CDM) model.  We emphasize that the model presented here is only a \it toy construction
of the late Universe\rm. The model is not intended to provide a consistent cosmology from the time of Big-Bang Nucleosynthesis (BBN) until today. It is 
unlikely that the precise model presented here is compatible with solar system experiments and the tight constraints on the time variation of Newton's constant. However, the model \it does \rm provide an example of how the postulated existence of a minimal curvature scale in gravity can provide a new mechanism to 
generate cosmological acceleration of the late Universe.  Furthermore, the model may capture features of a possibly more fundamental theory that admits a minimal curvature scale~\footnote{An interesting example of a minimal curvature scale occurs in a certain classical limit of quantum gravity~\cite{Schuller:2004nn,Schuller:2004rn}.}.

In Section~\ref{sec:MC}, we describe the
minimal curvature construction, first by using a toy example and then by using a class of modified Brans-Dicke theories.
We solve the equations of motion for this example and demonstrate how the Universe evolves from a matter dominated phase to
an accelerating period as the curvature approaches its minimal value. In Section~\ref{sec:comp}, we compare the MC model with $\Lambda$CDM and 
to the supernovae (SNeIa) ``gold"  sample of~\cite{Riess:2004nr}.
Finally, we comment on the possibility of constructing more realistic models that satisfy the limiting curvature hypothesis and offer our conclusions and speculations in Section~\ref{sec:conclusions}. In Appendix A, we provide a detailed analysis of the vacuum MC theory. In Appendix B, we construct an Einstein frame description of the vacuum theory and compare it to the MC vacuum.

\section{The Minimal Curvature Construction}
\label{sec:MC}
Our goal is to construct theories in which a certain physical quantity is bounded from below. Before
leaping directly into our model, it is instructive to consider an example of how a similar
effect may be achieved in a simpler theory - the bounding of velocities from above in Special Relativity by the
speed of light~\cite{Brandenberger:1993ef}.

The Newtonian action for a free particle of mass $m$ in motion is
\be
\label{old}
S = \int dt \,  \half \, m \,\dot x^2 \ .
\ee
In this classical theory the velocity of the particle is \emph{without bound}.
Now let us implement one of the fundamental consequences of Special Relativity:
To ensure that the speed of this particle is \emph{limited} by the speed of light 
we introduce a field $\p(t)$ which couples to the quantity in the action that we want to bound ($\dot x^2$) and has a ``potential" $U(\p)$.  The resulting action is
\be\label{newa}
S = m \int dt \,  \left[ \,\half \,\dot x^2 + \p \dot x^2 -U(\p) \, \right] \ .
\ee
The variational equation with respect to $\p$ 
\be\label{bit}
\dot x^2 = \frac{\d U}{\d \p} \ ,
\ee
ensures that $\dot x$ is bounded, provided $\d  U/ \d \p$ is bounded. Note the absence of a kinetic term for $\varphi$ in the
action, and hence, the reason the word \it potential \rm appears in quotes above. In order to obtain the 
correct Newtonian limit for small $\dot x$ and small $\p$ we take $U(\p)$ proportional to $\p^2$.
In the Newtonian limit the action (\ref{newa}) reduces to (\ref{old}). A simple potential satisfying the above asymptotics is
\be
U(\p) = \frac{2 \p^2}{1 + 2\p} \ .
\ee
Integrating out $\p$ yields (up to an irrelevant constant) the action for relativistic particle motion:
\be
S_{SR} = m \int dt \, \sqrt{1-\dot x^2}
\,.
\ee
The above model provides a powerful example of how a toy construction based on a fundamental principle -- the existence of a  universal ``speed limit" -- 
can capture features of a more fundamental theory. 

We now use a similar construction to model the existence of a \emph{minimal} curvature (MC) scale in gravity. 
Because we are interested in late
time cosmology, we need only be concerned with bounding one curvature invariant, the Ricci scalar
$R$. In direct analogy with our example from Special Relativity, we introduce a scalar field $\vp$ that couples to the quantity we wish
to bound, $R$, and a ``potential" function $V(\vp)$
\be
\label{act2}
S_{MC} = \int d^4 \! x \, \sqrt{-g} \,\left( \frac{\mpl^{2}}{2} R\, -\, \g \vp R \,+\, V(\vp) \right)\ ,
\ee
where $\mpl \equiv (8 \pi G)^{-1/2}$ is the reduced Planck mass and both $\vp$ and $\g$ posses dimensions of mass, $[\vp] = [\g]=M$. 
The vacuum theory is equivalent to a Brans-Dicke theory with Brans-Dicke parameter $\omega_{BD} =0$. This is seen by re-writing 
the action (\ref{act2}) 
in terms of a new field 
\be
\label{newp}
\Phi =  \left( \frac{\mpl^2} {2} - \g \vp \right)
\,,
\ee
 so that the action becomes
\be
\label{actbd}
S_{BD} = \int d^4 \! x \, \sqrt{-g} \,\left( \Phi R \,+\, V(\Phi) \right) 
\,.
\ee

Ordinarily, such a theory can be re-cast as a purely gravitational theory with Lagrangian 
$\mathcal{L} = \sqrt{-g} f(R)$ (see, e.g.~\cite{Magnano:bd}); however, this is not possible for
all forms of the potential $V$.

For reasons that will become clear shortly, we allow the field $\vp$ to couple non-minimally to matter.
The non-minimal coupling yields the matter action,
\be
\label{actmat}
S_M = \int d^4 x\, \sqrt{-g}\, {\mathcal{L}}_M(\psi_i, \vp, f(\varphi/M_{pl}) \bar\psi \psi) \,,
\ee
where ${\mathcal{L}}_M$ is the Lagrangian made up of whatever matter fields $\psi_i$ are in the theory. In the case were $\psi$ represents a 
dark matter Dirac spinor, the field $\vp$ couples
non-minimally \it only to the dark matter sector \rm and need not couple to baryons. In this case it is possible to avoid constraints on such a coupling 
from solar-system and table-top  tests of gravity. In string theory, non-perturbative string loop effects do not generically lead to universal couplings, 
allowing the possibility that the dilaton decouples more slowly from dark matter than ordinary matter (see, e.g.~
\cite{Damour:1990tw,Bean:2000zm, Bean:2001ys, Biswas:2004be, Biswas:2005wy, Bean:2005ru,Das:2005yj}). 

This coupling can be
used to address
the coincidence problem, since the acceleration is triggered by the coupling to matter.
For the purposes of our toy construction, we do not distinguish
between baryonic and dark matter in the remainder of our discussion.

Note that in the case of $f(R)$ theories which are conformally identical to models of quintessence in which matter is coupled to dark energy with a large coupling, this strong coupling induces a cosmological evolution radically different from standard 
cosmology~\cite{Amendola:2006kh}. Similar difficulties may arise in the model presented here, however, we have yet to investigate this issue.

The matter stress-energy tensor is given by
\be
T_{\m \n} \equiv - \frac{2}{\sqrt{-g}} \frac{\delta {\mathcal{L}_M(\psi_i, \vp)}}{\delta g^{\m\n}} \,.
\ee
We assume a perfect fluid
\be
T_{\m\n} = (\rho_M + P_M) u_{\m}u_{\n}+P_M g_{\m\n}
\,,
\ee
where $u^{\a}$ is the fluid rest-frame four-velocity, and the energy density $\rho_M$ and pressure $P_M$ are related by the equation of state $P_M =w \rho_M$. 
Because we are focusing on the late Universe we shall ignore the presence of radiation and consider only a matter density
$\rho_M$, which redshifts with expansion in the usual manner (with the exception of the non-trivial $\vp$ dependence)
\be
\rho_M \propto \frac{f(\vp/\mpl)}{a^3} \,.
\ee
In the above, $f$ is a function describing the non-minimal coupling of the field $\vp$ to the matter Lagrangian. Such couplings in the context of ordinary 
scalar-Einstein gravity were studied in~\cite{Das:2005yj} 
where it was found that this coupling can be made consistent with all known current observations, the tightest constraint coming from
estimates of the matter density at various redshifts. This coupling plays a critical role in Modified Source Gravity, introduced
in~\cite{Carroll:2006jn}.

Variation of the total action $S_{tot} = S_{MC} + S_M$ with respect to the metric tensor, $\delta S_{tot}/\delta g^{\m\n}=0$ yields the modified Einstein
equations
\bea
\label{eomg}
(1- \a \vp) R_{\m\n} &-&  \half(1- \a \vp) \, R\, g_{\m\n}  \nonumber \\
&+& \a \nabla_{\m} \nabla_{\n}\vp - \a g_{\m\n} \Box \vp -   V\, g_{\m\n} = 8 \pi G \, T_{\m\n}
\eea
where $\opensquare  \equiv g^{\m\n} \nabla_\m \nabla_\n$ and we have introduced $\a \equiv 2 \g/\mpl^2$.
Variation with respect to the field $\vp$ gives
\be
\label{eomp}
\g R =  \frac{\d V }{\d \vp}   +  \frac{1}{f_0} \frac{\d f(\vp/\mpl) }{\d \vp} {\rho_M}\,.
\ee
Equation (\ref{eomp}) is the key to imposing the limiting curvature construction. It is clear that the curvature $R$ will remain bounded 
and approach a constant curvature at late times ($t \rightarrow \infty$) as long as $ V' + f' \r_M \rightarrow$ constant; this is the essence of the construction.
We assume a flat ($k=0$) Friedmann-Robertson-Walker metric 
\be
ds^{2}=-dt^{2}+a^{2}(t)\left\{dr^{2} + r^{2} d\Omega^{2}\right\} \ ,
\label{frwm}
\ee
where $a(t)$ is the scale factor of the Universe and $d\Omega$ is the line-element of the unit $2-$sphere.
Defining the Hubble parameter by $H \equiv \dot a/a$ and substituting the metric ansatz (\ref{frwm}) into (\ref{eomg}) and (\ref{eomp}) gives the 
generalized Friedmann equation (the $00$-component of (\ref{eomg})) and the equation of motion for $\vp$:
\bea
\label{eom1}
3\mpl^2 H^2 - 6 \g \vp H^2 - 6 \g H \dot \vp + V(\vp) = \frac{f(\vp/\mpl) \r_M^{(0)}}{f_0 a^3}\,, &\quad& \nonumber \\
6\g(2H^{2} + \dot H) - V'(\vp)  = f'\left(\frac{\vp}{\mpl}\right) \frac{ \r_M^{(0)}} {f_0 a^3}\ ,
\eea
where a prime denotes differentiation with respect to $\vp$, $f_0 \equiv f(\vp_0/\mpl)$, $\vp_0$ and $ \r_M^{(0)}$ are the values of $\vp$ and $ \r_M$ today.

By considering the asymptotics of our cosmology at early and late times, we can constrain the
forms of the functions $V(\vp)$ and $f(\vp)$. We require that the effective Newton's constant for
our theory remain positive definite so that gravity is always attractive. This imposes a constraint on $\Phi \ge 0, \; \forall \; t$ in \eq{actbd}.
There are rather strong constrains on the time variation of Newton's constant from the period of nucleosynthesis until today (roughly, $\left|\dot G/G\right| < \mathcal{O}(10^{-10} - 10^{-13}\, yr^{-1})$~\cite{Uzan:2002vq}).
For the time being, we will allow ourselves to ignore this constraint in order to produce a toy model capable of
realizing the MC conjecture. Furthermore, because we are only interested
in the behavior of the Universe from the matter dominated epoch until today, we have ignored the presence of radiation. The absolute earliest
our theory is valid is up to the period of equal matter and radiation domination $t_{eq}$. 
For specificity, by \it early times \rm we refer to times near the time of photon decoupling at a redshift of $z \simeq 1100$, during which
the Universe is typically already well into the matter dominated regime.

At late times (and low curvatures) we want to bound the Ricci scalar $R$ from below. This will constitute a successful example of a model obeying the minimal curvature hypothesis.
To bound the curvature we use \eq{eomp}. It is clear that the curvature $R$ will remain bounded if we bound $V' + f' \r_M$, where the prime denotes differentiation with respect to $\vp$.
Hence,
we require $\g^{-1} (V' + f' \r_M) \rightarrow \mathcal{R}$ at late times, where we denote the hypothesized minimal curvature scale by $\mathcal{R}$. 
We anticipate that, by construction, there
will be a late-time attractor that is a constant curvature space-time with $R=\mathcal{R}$.  This attractor is not an
actual solution to the equations of motion.
The above considerations restrict the functional forms of potential $V$ and the non-minimal coupling function $f$. 
Integration singles out a class of theories that must obey $V(\vp) + f(\vp) \r_M \rightarrow \g \mathcal{R} \vp$ as $t \rightarrow \infty $.
The simplest forms for $V$ and $f$ obeying the above constraint are the linear functions
\be\label{vandf}
V(\vp) = \m^3 \vp \,, \qquad f(\vp) = \frac{\vp}{\mpl}
\,,
\ee
where $\m$ is, in principle, another free parameter with dimensions of mass. However, we will take $\m = \g$, so that $\mathcal{R} = \g^2$, and $\g$ is the only free parameter in the theory~\footnote{It is interesting to note that if we do not take $\m=\g$ and then take the
limit that $\g \rightarrow 0$ the theory resembles the action of the \it modified source gravity \rm models studied in~\cite{Carroll:2006jn}.}. We now rescale time $t \equiv t H_0$ so that today $t_0 = 1$ and introduce the following dimensionless quantities
\be\label{dimen}
a \equiv \frac{a}{a_0}, \; H \equiv \frac{H}{H_0} \; \varphi \equiv \frac{\varphi}{M_{pl}}, \; \gamma \equiv \frac{\gamma}{M_{pl}}
\ee
\be\label{dimen2}
\Omega_M \equiv \frac{\rho_M}{\rho_c} = \frac{\rho_M}{3H^2 \mpl^2}, \; R = \frac{R}{H_0^2},
\ee
and take $a_0 =1$, where $a_0$ is the value of the scale factor today.

In terms of the dimensionless quantities the EOM become
\bea
\label{eomf}
H^2 - 2 \g \vp H^2 - 2 \g H \dot \vp + \frac{V(\vp)}{3} = \frac{f(\vp) \Omega_M^{(0)} }{f_0 a^3}\,, &\quad&  \\
\g(2H^{2} + \dot H) - \frac{V'(\vp)}{6}  = f'\left({\vp}\right) \frac{\Omega_M^{(0)}} {2f_0 a^3}\label{rdshift}\ ,
\eea
The successful implementation of the minimal curvature hypothesis is now apparent. Recasting equation (\ref{rdshift}) in terms of the curvature scalar
$R = 6 (2 H^2 + \dot H)$, and substituting in our choices for $V$ and $f$ (\ref{vandf}):
\be\label{simple}
R = \g^2 + 3 \frac{\Omega_M^{(0)}}{a^3}
\,.
\ee
We see that, as the Universe expands and the matter term dilutes, we asymptotically approach the minimal value of the
curvature $R= \mathcal R = \g^2$. 

It is both interesting and surprising that the solution to Eq.~(\ref{simple}) reduces to the simple case of 
$\Lambda$CDM
plus an arbitrary amount of  \emph{dark radiation} which may have either positive or negative effective energy density.  
Most notably, this model arises in Randall-Sundrum brane cosmology which has been extensively studied in the literature~\footnote{For some
reviews see, e.g.~\cite{Easson:2000mj,Langlois:2002bb}.}. The Friedmann equation derived from the Randall-Sundrum model for a flat Universe is
\be\label{bwf}
H^2 =  \frac{\rho}{3 \mpl^2} + \frac{\rho^2}{ 36 M_5^6} + \frac{\mathcal C}{a^4}
\,,
\ee
where $M_5$ is the five dimensional Planck mass and $\mathcal C a^{-4}$ is the so called dark radiation term, since it scales like radiation, but it's origins are purely gravitational and it does not interact with standard matter~\cite{Binetruy:1999hy}. At low energies (when the energy density is much less than the critical brane tension), the $\rho^2$ term can be safely neglected. 
The main observational restrictions on the dark radiation term come from the acoustic scale at recombination~(see, e.g.~\cite{Wang:2006ts}),
and from the amount of total growth of density perturbations in the non-relativistic matter component from the time of equal matter
and radiation until the present day~\cite{Doran:2006kp,Linder:2006ud}. As a result, the density of dark radiation cannot be
significantly larger than the present CMB energy density~\footnote{Note, the constraints from BBN are even stronger~\cite{Langlois:2005nd}.}.

Making use of (\ref{dimen}) and  (\ref{dimen2}), Eq.~(\ref{bwf}) may be recast (neglecting the $\rho^2$ term) as
\be\label{hbw}
H^2 = \frac{\Omega_M}{a^3} +  \frac{\Omega_R}{a^4}  + \frac{\Omega_\Lambda}{3}
\,,
\ee
where we have included a cosmological constant term $\Omega_\Lambda = \Lambda/3\mpl^2 H^2$, and 
$\Omega_R = \Omega_\mathcal C+ \Omega_r$ includes contributions from both the dark and ordinary radiation.
From (\ref{hbw}) we find
\be
\dot H = \frac{-3 \Omega_M}{ 2 a^3} - \frac{2\Omega_R}{a^4}
\,.
\ee
Constructing the Ricci Scalar $R=6(2H^2 + \dot H)$ from the above expressions yields Eq.~(\ref{simple}), with $\g^2 = 4 \Omega_\Lambda$.

In \ref{app:vac}, we provide a detailed analysis of the vacuum MC equations with
$\Omega_M =0$. Although the presence of matter plays an important role in our minimal curvature construction, an analysis
of the vacuum theory provides valuable insight into the solutions we are interested in studying. In \ref{app:einst},
we transform the vacuum MC theory into an Einstein frame and relate quantities of physical interest in both frames.

\begin{figure}[ht]
\begin{centering}
\includegraphics[width=5.0 in,clip]{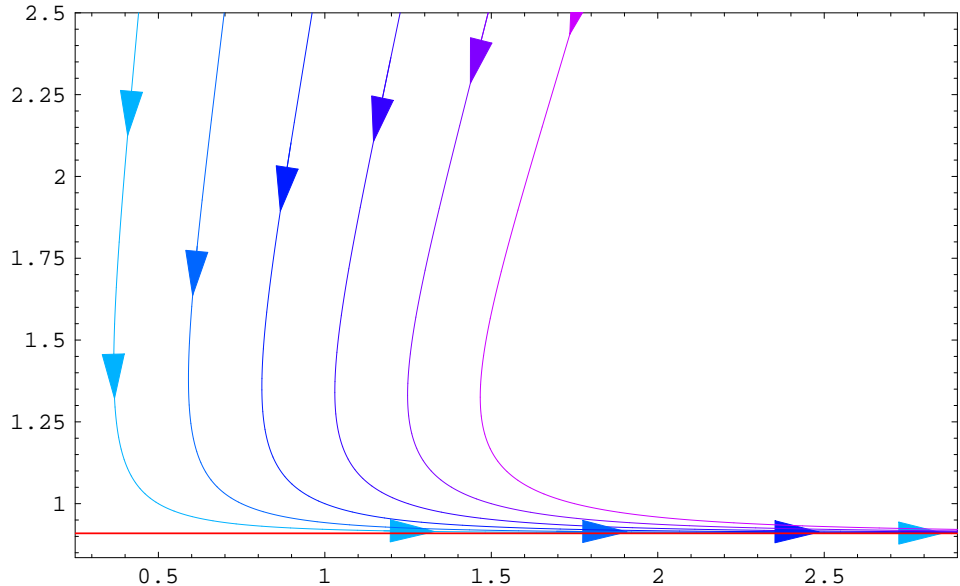}
\caption{Phase space of $\vp$ vs. $H$. All solutions asymptotically approach
a late-time accelerating phase with constant $H$ at the minimal curvature $\mathcal R$ denoted by the red line.}
\label{phase1.eps}
\end{centering}
\end{figure}

For general solutions to the equations of motion~(\ref{eomf}) and (\ref{rdshift}) with functions~(\ref{vandf})
we solve for the Hubble parameter $H(t)$ and scalar field $\vp(t)$. We plot the relevant portion of the
$\vp\!-\!H$ phase space in Fig.~\ref{phase1.eps}. 

To solve the equation we integrate from the past to today $(H_0 =1)$ and then from today
into the future and patch the solutions together. In the plots we take the conditions $H_0=1.0$,
$\g=3.15$, $\Omega_M = 0.25$ and let the values of $\vp_0$ vary. At late times, the solutions approach the constant $H$ attractor
when the Universe is well into the accelerating epoch.

\section{Comparison with $\Lambda$CDM}\label{sec:comp}
We now compare our model with $\Lambda$CDM. To do so, we must enter reasonable initial conditions into our numerical study
and solve the equations of motion (\ref{eomf}),  (\ref{rdshift}) together with (\ref{vandf}) and the equation $\dot a = H a$. Let us begin
by considering the value of the minimal curvature. Physically the minimal curvature corresponds to an emptying of the matter in the
Universe due to cosmological expansion. In our model the value of the minimal curvature is approached asymptotically. Today, the value of the 
curvature is given by
\be\label{Rprox}
R_0 = 6 (2 H_0^2 + \dot H_0) \approx 12 H_0^2
\,.
\ee
Because we observe a significant amount a matter in the Universe today, we know that we have not yet reached the minimal curvature. However,
because we are accelerating, we know that we are \it near \rm the minimal curvature (i.e.~the first and second terms on the
right hand side of \eq{simple} must be comparable). Hence, from \eq{Rprox}, we expect the value of the minimal curvature
$\mathcal R$ to be close to but less than $\mathcal R < 12 H_0^2$. Therefore, in terms of our dimensionless quantities (\ref{dimen}),
$\mathcal R < 12$ and the free parameter in our theory $\g < \sqrt{12} \approx 3.4641$. For the solutions considered below,
we choose a value of $\g=3.15$ meeting the above requirements and that follows the $\Lambda$CDM model to our satisfaction for our toy construction
(i.e. a value leading to a matter dominated cosmology followed by a ``jerk" near a redshift of $z \simeq .5$ into an accelerating phase). 

In the action (\ref{act2}), the effective Newton's constant is $G_{N_{eff}}$ is given by
\be
(16 \pi G_{N_{eff}})^{-1} = \frac{\mpl^2}{2} - \g \vp
\,.
\ee
To ensure that the effective Newton's constant remains positive definite over the history of the Universe ( $G_{N_{eff}} > 0, \, \forall \, t$) we must have $\g \vp <.5$.
We are almost in a position to compare our model both with $\Lambda$CDM, which has the  Friedmann equation
\be\label{lcdm}
H = \sqrt{ \frac{\Omega_M^{(0)}}{a^3} + (1 - \Omega_M^{(0)})}
\,,
\ee
and with the observational data provided by the SNeIa gold sample. To make a comparison with the observational data we
require an understanding of the luminosity distance formula in the context of modified gravity models.

An important consideration arises when using the formula for the luminosity distance in theories of the form:
\be\label{modact}
S = \int d^4 x\, \sqrt{-g}\, \left[ A(\varphi) \frac{R}{2} + B(\varphi) (\partial_\mu \varphi)^2 + V(\varphi) +  \sum_i C_i(\varphi) \mathcal L_i \right] 
\,,
\ee
where the $\mathcal L_i$ represent the different types of possible matter Lagrangians present.
Such a theory arises as the low-energy effective action for the massless modes of dilaton gravity in string theory, and our model is an action of 
just this sort~\cite{Damour:1994ya,Damour:1994zq}; albeit, with an unusual choice for the functions $A,\, B,\, V,\, C$. As we have already discussed, these theories typically lead to time variation in Newton's 
gravitational constant. The time variation can affect the way one should compare the theory to observations~\cite{Acquaviva:2004ti}. In particular,
the time-evolution can alter the basic physics of supernovae. For example, the time variation in $G_N$ leads to different values of the Chandrasekhar mass at different
epochs, and hence, a supernova's peak luminosity will vary depending on when the supernova exploded. This makes treating the supernovae Ia as standard
candles difficult~\cite{Amendola:1999vu, Garcia-Berro:1999bq,Riazuelo:2001mg,Gaztanaga:2001fh,Nesseris:2006jc,Gannouji:2006jm}. 
Specifically, the peak luminosity of SNeIa is proportional to the mass of nickel synthesized which is a fixed fraction of the Chandrasekhar mass $M_{Ch} \sim G^{-3/2}$. Hence, the luminosity peak of SNeIa varies as $L \sim G^{-3/2}$ and the corresponding absolute magnitude evolves as
\be
M = M_0 + \frac{15}{4} \log{\frac{G(z)}{G_0}}
\,,
\ee
where the subscript zero indicates the local values of the quantities. Therefore, the magnitude-redshift relation of SNeIa in modified
gravity theories of the type given by (\ref{modact}) is related to the luminosity distance via~\cite{Riazuelo:2001mg,Nesseris:2006jc}:
\be
m(z) = M_0 + 5 \log{d_L(z)} + \frac{15}{4} \log{\frac{G(z)}{G_0}}
\,.
\ee
Even if gravitational physics is described by some theory other than General Relativity the standard formula for the luminosity distance 
applies as long as one is considering a metric theory of gravity~\cite{Shapiro:2005nz}:
\be\label{lumd}
d_L(z;  H(z), H_0)  = \frac{(1+z)}{H_0}  \int_0^z \frac{dz'}{H(z')}
\,.
\ee

For $\Lambda$CDM, the Luminosity distance (\ref{lumd})  can be written~\cite{Perlmutter:1996ds}:
\bea\label{lumlcdm}
d_L(z;  \Omega_M, \Omega_\Lambda, H_0)   =&& \nonumber \\
 \frac{c (1+z)}{H_0 \sqrt{|\k|}} \, {\mathcal S} \left( \sqrt{|\k|} \int_0^z \left[ (1 + z')^2 (1 + \Omega_M z') -z'(2+z')\Omega_\Lambda \right]^{-\half} \right)
\eea
where, $\mathcal S(x) \equiv \sin(x)$ and $\k = 1-\Omega_{tot}$ for $\Omega_{tot} >1$ while $\mathcal S(x) \equiv \sinh(x)$ with $\k = 1 - \Omega_{tot}$
for $\Omega_{tot} <1$ while $\mathcal S(x) \equiv x$ and $\k=1$, for $\Omega_{tot} =1$. Here and throughout, $\Omega_{tot} = \Omega_M + \Omega_\Lambda$.

\subsection{Numerical Analysis}
Given the above considerations we take the following values of parameters today:
\be\label{initc}
a_0 =1.0, \quad H_0 = 1.0,\quad \vp_0 = 0.1, \quad \g = 3.15, \quad \Omega_M^{(0)} = 0.25
\,.
\ee
We then integrate our equations from the past to today and then from today into the future and patch the solutions together.
While we do not provide an exhaustive study of the parameter space for the MC model, the parameters given above provide a
successful example of the construction which fulfills our rather modest goals~\footnote{An exhaustive study of the phase space of the vacuum theory
is supplied in \ref{app:vac}.}. It is quite possible that the parameters~(\ref{initc}) can be tuned
to achieve an even better agreement with $\Lambda$CDM.

\subsection{Results}
\begin{figure}[ht]
\begin{centering}
\includegraphics[width=\columnwidth]{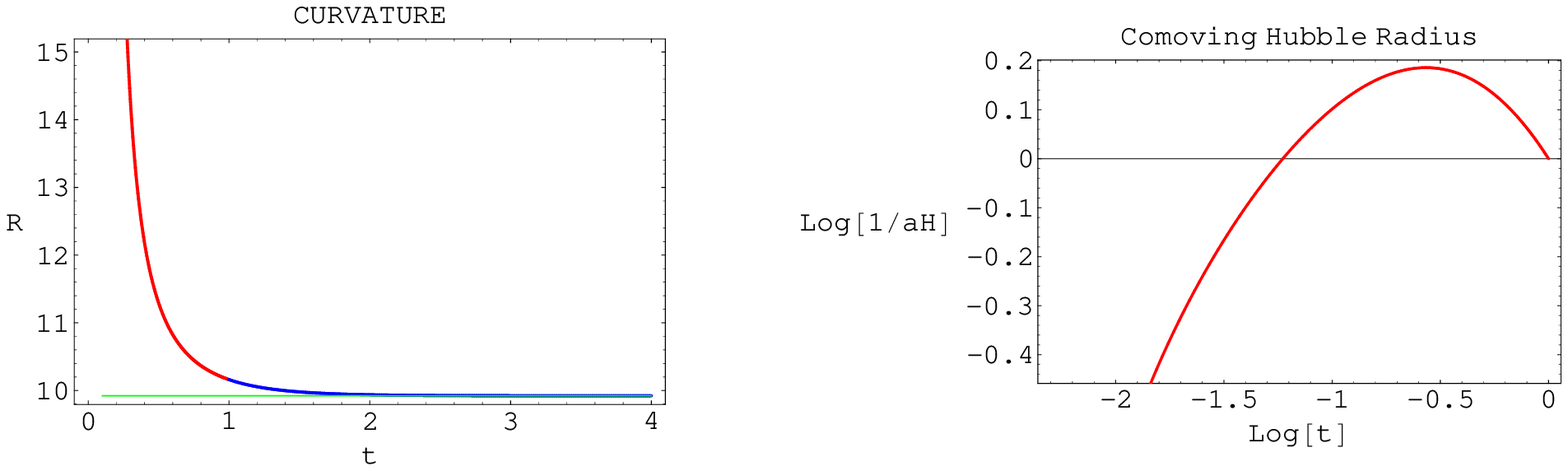}
\caption{The Ricci scalar (left) and co-moving Hubble radius (right) as functions of cosmic time $t$. The curvature decreases from the matter-dominate past (red curve) to the constant minimal value
   at $R=\g^2$ indicated by the green line. Today, $t_0=1$ and the blue curve indicates the future evolution of the curvature. While the Universe is accelerating the co-moving Hubble radius decreases.}
\label{crvandcmh2.eps}
\end{centering}
\end{figure}

The history and future of the curvature $R$ together with the co-moving Hubble radius are plotted in Fig.~\ref{crvandcmh2.eps}.

In the Figure the past history of the curvature is plotted in red while the future is plotted in blue. Today we are at the value $t =1$. The minimal curvature is given
by the green line at $R=\mathcal{R} = \g^2$. As expected, the curvature is large in the past when 
the Universe is matter dominated and then decreases, approaching the accelerating late-time de-Sitter attractor with constant minimal curvature $R=\mathcal{R}$. The second plot in the Figure shows the evolution of the co-moving Hubble radius $ H^{-1}/a$. The 
phenomenologically desired cosmological
transition from matter domination to late-time acceleration of the Universe is clearly indicated by the decreasing of the co-moving Hubble radius, when
\be 
\frac{d}{dt} \frac{H^{-1}}{a} < 0
\,.
\ee
In some modified gravity models it can be difficult to achieve this transition (see, e.g.~\cite{Easson:2005ax,Punzi:2006bv}).

\begin{figure}[ht]
\begin{centering}
\includegraphics[width=\columnwidth]{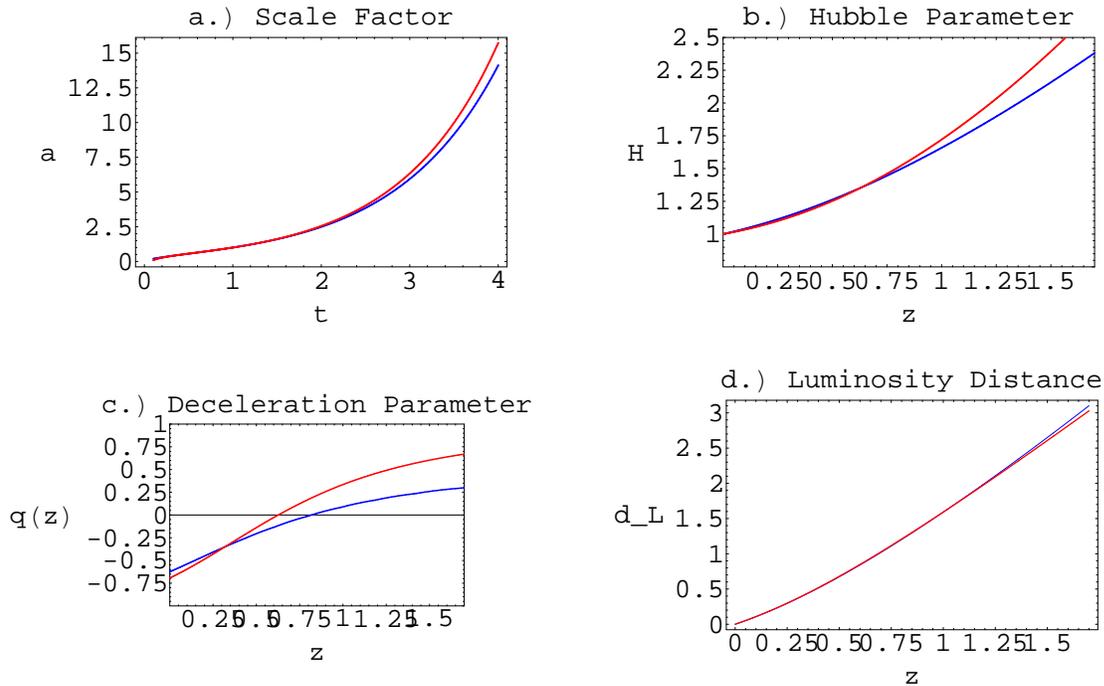}
\caption{A comparison of the cosmological evolution of the MC model with $\Lambda$CDM. Quantities
   plotted in red are the MC model while those in blue are for $\Lambda$CDM. Today we are at the values $t=1$, $z=0$. Plot a.) shows the time evolution of the scale factor. Plot b.) 
   shows the Hubble parameter as a function of redshift. Plot c.) compares the deceleration parameters as a function of redshift. Plot d.) shows
   the luminosity distance, $d_L$, as a function of redshift. The MC model is clearly a good 
   fit with $\Lambda$CDM.}
\label{vslcdm3.eps}
\end{centering}
\end{figure}

\begin{figure}[ht]
\begin{centering}
\includegraphics[width=5in,clip]{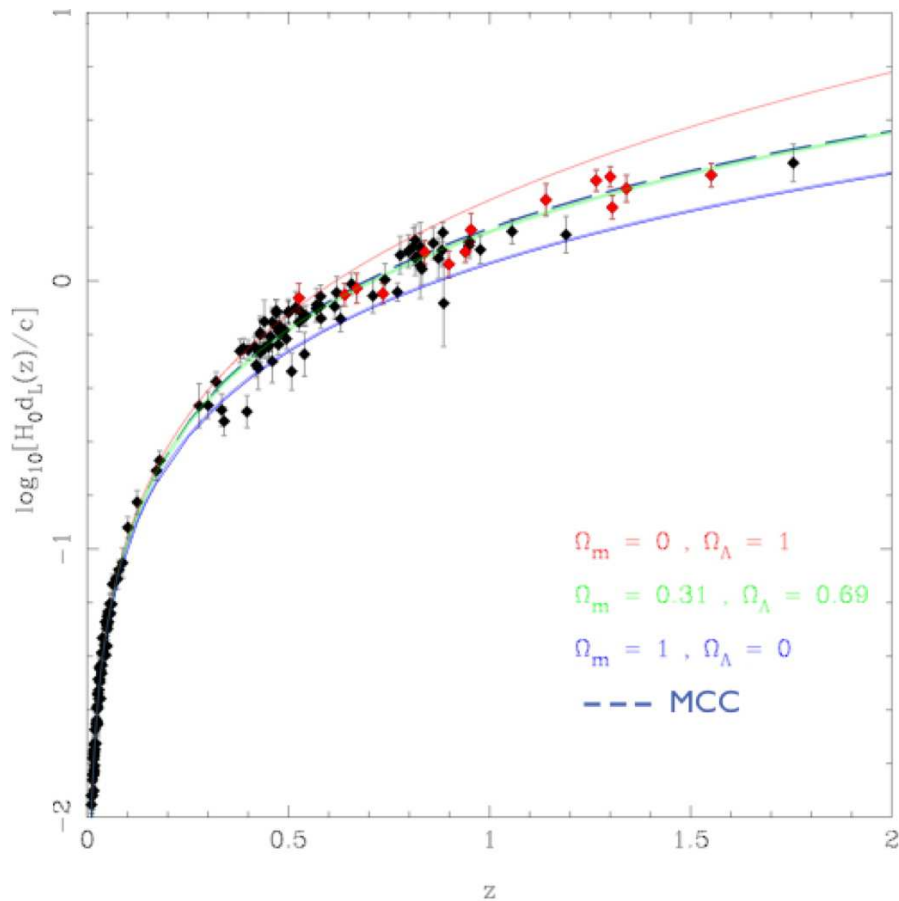}
\caption{Comparison of the MC and various $\Lambda$CDM models with observational data. The supernovae data
   points are plotted with error bars and the data is taken from~\cite{Riess:2004nr}. The luminosity distance $d_L$ for the MC model is plotted by the dashed (blue) curve. The
   various theoretical predictions for $\Lambda$CDM are represented by the solid curves and were examined in~\cite{Choudhury:2003tj}.}
\label{mcfit4.eps}
\end{centering}
\end{figure}

The majority of our results are presented in Figures \ref{vslcdm3.eps} and \ref{mcfit4.eps}.
In Figure \ref{vslcdm3.eps} we plot the scale factor $a$ as a function of cosmic time $t$, the Hubble parameter $H$ as function of redshift $z$, where
\be
z + 1 \equiv \frac{a_0}{a}\,,
\ee
the deceleration parameter
\be
q(z) = - \frac{\ddot a}{a H^2}
\,,
\ee
and the luminosity distance, $d_L$.

\newpage

For the particular set of parameters considered, the scale factor of our model differs from $\Lambda$CDM most strongly in the
far future, although the difference in the expansion rates of the MC construction with $\Lambda$CDM is apparent in the plot of the
Hubble parameters at high redshifts.
The transition from matter domination to acceleration (the jerk) occurs at a slightly lower redshift than $\Lambda$CDM. There is only a slight
difference in $d_L$ that occurs at high redshifts ($z>1$), although the difference is not significant enough to distinguish our model
from $\Lambda$CDM using only the current supernova data. 
\newpage

In Figure \ref{mcfit4.eps}, we compare the luminosity distance predicted by our model with several versions of $\Lambda$CDM and with the observational data. 
The luminosity distances are plotted out to a redshift of $z=2$ (the highest redshift supernova data is from $z = 1.76$). The theoretical predictions of the minimal
curvature construction and $\Lambda$CDM are compared
with the ``gold" supernovae sample of~\cite{Riess:2004nr}. The particular choice of $\Lambda$CDM models shown are from~\cite{Choudhury:2003tj}.
The luminosity distance for the minimal curvature construction is
denoted by the blue dashed line and fits the supernova data extremely well. The luminosity distance $d_L$ is virtually indistinguishable from $\Lambda$CDM with $\Omega_M^{(0)} = 0.25$ and $\Omega_\Lambda^{(0)} = 0.75$ (there is a small discrepancy at high redshift as can be seen from Figure \ref{vslcdm3.eps}). 

\section{Conclusions}
\label{sec:conclusions}
We have shown that a period of late-time cosmic acceleration can follow directly from a simple
minimal curvature conjecture (MCC). The model fits the SNeIa data exceptionally well. While the specific formulation considered here is only a toy construction, unlikely to be compatible with constraints from solar system and table top test of the equivalence principle, it may capture phenomenologically interesting features of
a more fundamental theory that admits a limiting minimal curvature. Furthermore, the construction successfully demonstrates the possibility that a 
new fundamental physical principle  may ultimately be responsible for the recent period of cosmological acceleration.
It is possible that experimentally viable models based on the minimal
curvature conjecture exist. The search for such models within the context of scalar-Gauss-Bonnet gravity is currently underway.
Despite the tight theoretical and experimental constraints on scalar-Gauss-Bonnet cosmologies~\cite{Nojiri:2005jg,Cognola:2006eg,DeFelice:2006pg,Calcagni:2006ye,Koivisto:2006xf}, we remain optimistic that an experimentally and theoretically viable model
based on the minimal curvature construction can be discovered.

\section*{Acknowledgments}  

It is a pleasure to thank R. Brandenberger, R. Gregory, V. Jejjala, I. Moss, R. Myers and T. Underwood for helpful discussions.
I am especially grateful to M.~Trodden for numerous useful discussions over the course of this work.
This work is supported in part by PPARC and by the EU 6th Framework Marie Curie Research and Training network ``UniverseNet" (MRTN-CT-2006-035863).

\newpage
\appendix
\section{Probing the Vacuum Theory}\label{app:vac}
The basic mechanism used to implement the limiting curvature
construction requires a non-minimal coupling of $\vp$ to the matter Lagrangian. This coupling plays an important role in
the overall cosmological dynamics discussed above. Hence, an understanding of the vacuum theory is perhaps not as
fruitful as it would be in case of a theory with minimal coupling of $\vp$ to matter; however, once the Universe becomes sufficiently 
large and dilute the dynamics will resemble those
of the vacuum theory. As we shall see the vacuum MC theory is significantly richer than that of $\Lambda$CDM. The equations describing the vacuum are given by the equations (\ref{eom1}) with
$\Omega_M = 0$. The dimensionless forms are
\bea
\label{eomvac}
H^2 - 2 \g \vp H^2 - 2 \g H \dot \vp + \frac{V(\vp)}{3} =0\,, &\quad& \nonumber  \\
\g(2H^{2} + \dot H) - \frac{V'(\vp)}{6}  =0 \ ,
\eea
where we have re-introduced the parameter $\m$ from \eq{vandf} (previously set equal to $\g$) for completeness and the potential is 
$V(\vp) = \mu^3 \vp$. In the case of the vacuum we may solve exactly for $H(t)$. We focus on two types of solutions of particular interest. 
In the first, the Hubble parameter is given by
\be\label{htanh}
H(t) = \sqrt{\frac{1}{12}\mathcal R}  \, \tanh{\left\{  \sqrt{\frac{\mathcal R}{3}}  (t - 6\a \g)\right\}}
\,,
\ee
where$\mathcal R = \m^3/\g$ is the minimal curvature and $\a$ is a constant. In this case
\be
\dot H = \frac{\mathcal R}{6}  \, \cosh{\left\{  \sqrt{\frac{\mathcal R}{3}}  (t - 6\a \g)\right\}}
\,,
\ee
and the scale factor evolves as
\be
a(t) = a_0 \left[ \mbox{sech} \left\{ \sqrt{\frac{\mathcal R}{3}} (6\a \g - t) \right\} \right]^{\half}
\,.
\ee
The quantities mentioned above, along with the field $\vp$ are plotted in \fig{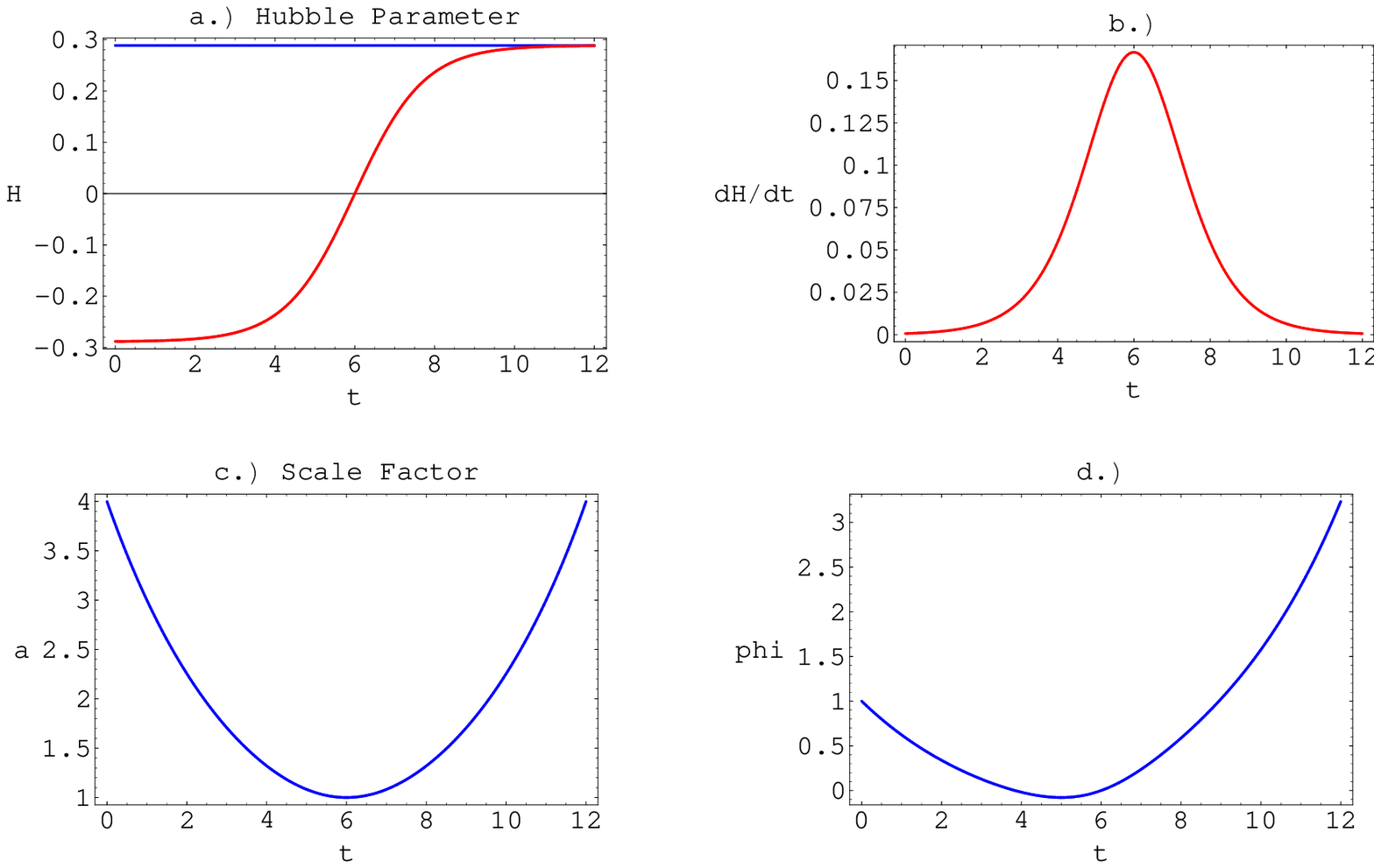}.
\begin{figure}[ht]
\begin{centering}
\includegraphics[width=\columnwidth,clip]{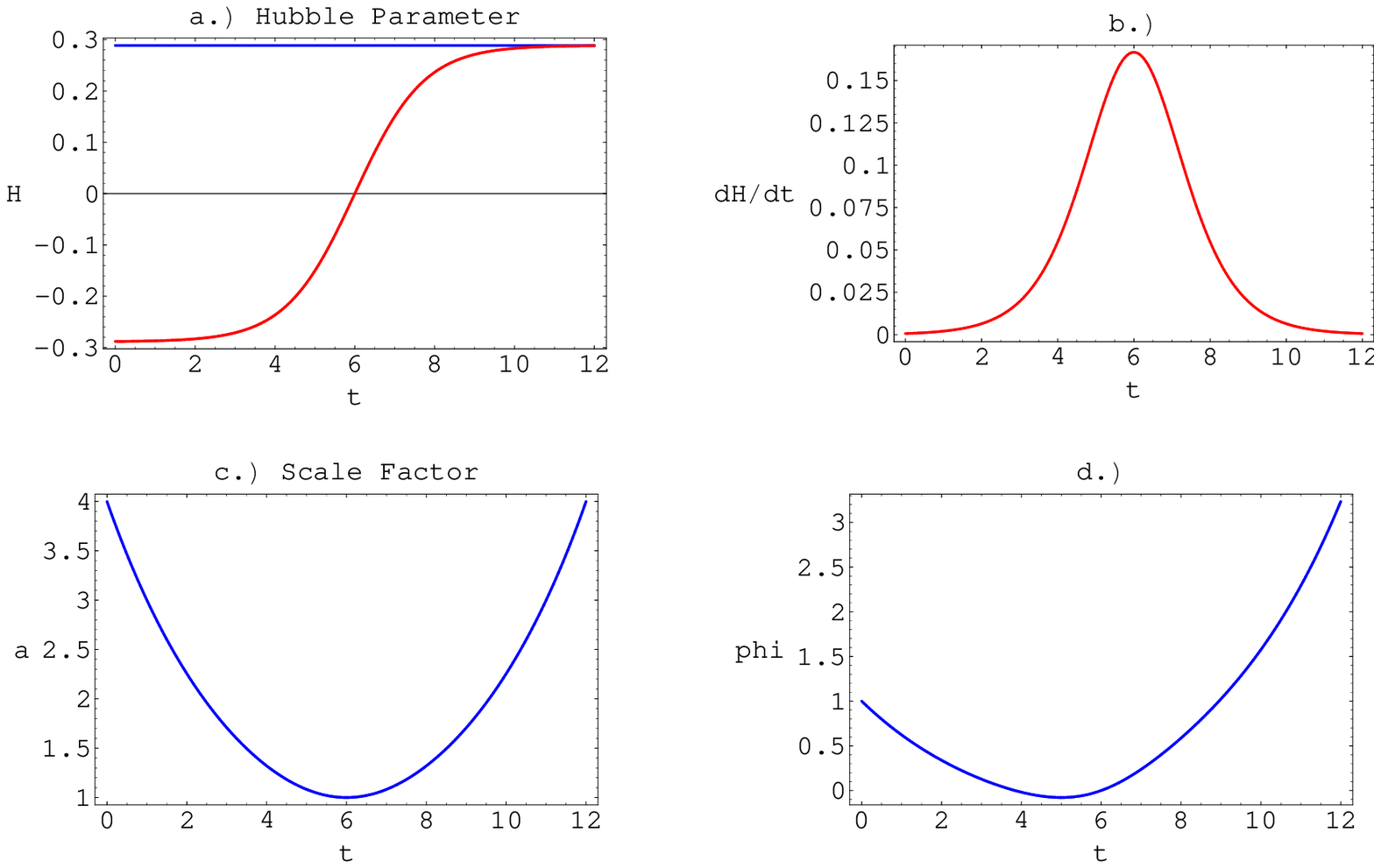}
\caption{Various quantities of interested plotted as functions of cosmic time $t$. In the plots we take $\mathcal R = \g = \m = \a = 1$.}
\label{vacplot.eps}
\end{centering}
\end{figure}
In this case the Universe undergoes a bounce as indicated in plots a.) and b.) in \fig{vacplot.eps}. After the bounce
the Hubble parameter is pulled up  to the value at the minimum curvature $H \rightarrow \sqrt{\mathcal R/12}$.
This is a de Sitter solution of the vacuum theory with 
\be\label{julia}
H_{dS} = \sqrt{\mathcal R/12}
\,, 
\ee
while $\vp$ continues to evolve
exponentially $\vp(t) + (2\g)^{-1} \propto \exp(\sqrt{\mathcal R/12})$.
Interestingly, $H$ and $\dot H$ evolve in such a way that the curvature $R = 6(2H + \dot H)$ is constant throughout the evolution
of the solution, fixed at the minimal value $R=\mathcal R$. This behavior is not surprising as the constraint on $R$ comes
from the second equation in the EOM (\ref{eomvac}).

\begin{figure}[ht]
\begin{centering}
\includegraphics[width=5 in,clip]{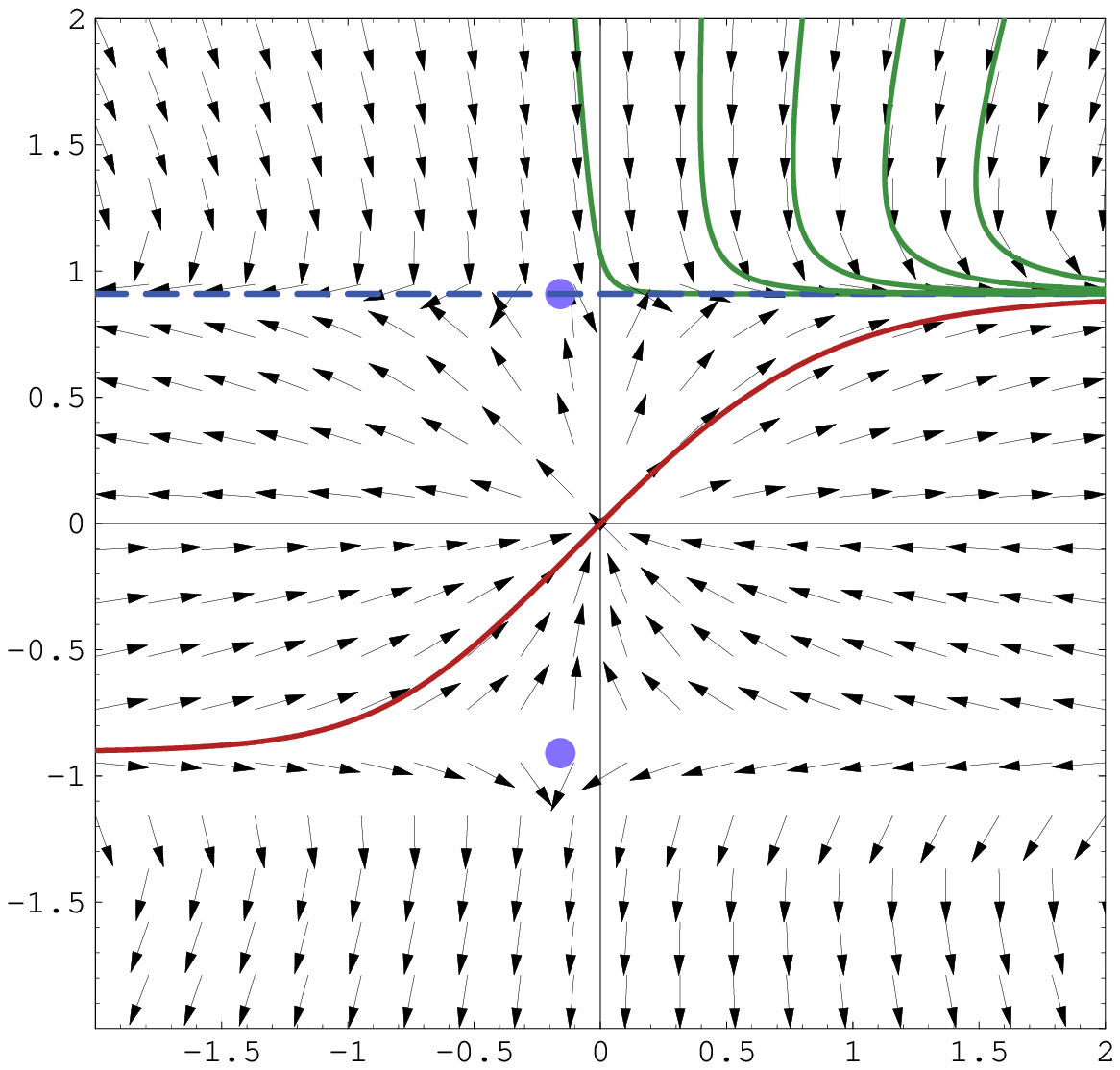}
\caption{The phase space of the vacuum theory. $H$ is plotted on the vertical axis and $\vp$ is plotted on the horizontal axis. The
vector field for solutions is drawn in black. Solutions of particular interests are plotted by the solid curves. The constant de Sitter value
of $H$ at the minimal curvature is indicated by the dashed (blue) line. The two unstable saddle equilibrium points are designated by
the purple dots. In the plot we take $\g = \m =3.15$.}
\label{vacphase.eps}
\end{centering}
\end{figure}

The second family of solutions are of greater relevance. They are the vacuum analogs of the solutions plotted in
\fig{phase1.eps} and discussed at the end of 
Section~\ref{sec:MC}.  These solution are pulled \it down \rm to the value of $H$ at the minimal curvature. The entire phase
space of solutions to the vacuum theory is shown in \fig{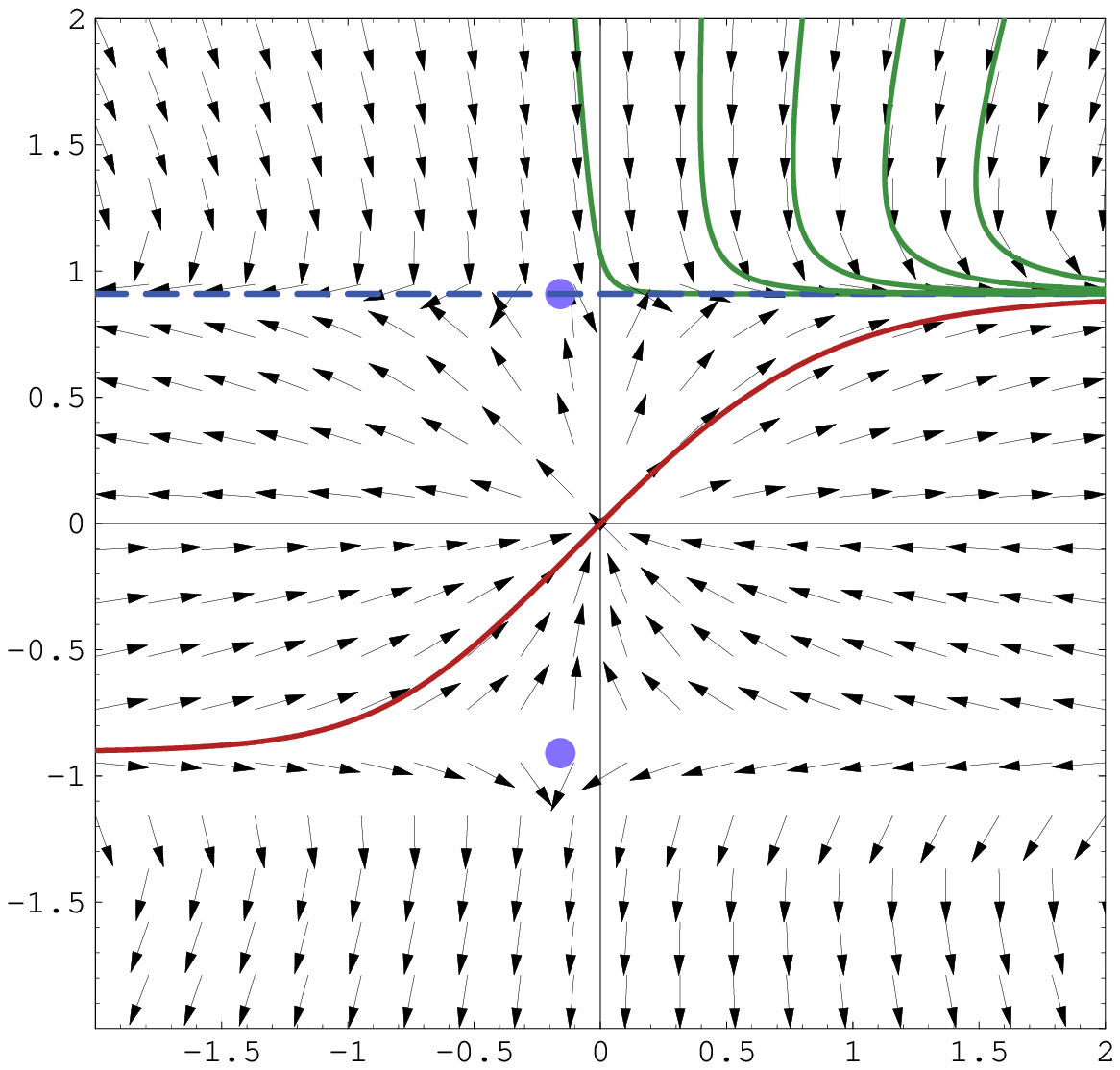}. Representative solutions of the two families of solutions
discussed above are plotted by the solid curves. The red curve is the solution given by \eq{htanh}. The second set of solutions
(relevant to Section~\ref{sec:MC}) are plotted in green in the upper right quadrant of the phase space. The system defined by the vacuum equations (\eq{eomvac})
has two unstable saddle equilibrium points at the values
\be\label{eqpts}
(\vp,H) = \left(-\frac{\mpl^2}{2\g}, \, \pm \sqrt{ \frac{\mathcal R}{12}}\right)
\,.
\ee
The equilibrium points are marked by the purple dots in \fig{vacphase.eps}.

\newpage
\section{An Einstein Frame Description}\label{app:einst}
It is not our intent to provide a complete Einstein frame analysis of the full MC system with matter; however,
it is instructive to consider the vacuum theory in an Einstein
frame. To move to the Einstein frame we begin with the Brans-Dicke frame defined by the action~(\ref{actbd}). Passage to the Einstein
frame is achieved via a conformal transformation of the form
\be\label{ctrans}
\tilde g_{\m\n} = \Omega^2 g_{\m\n}
\,,
\ee
where $\Omega^2$ is the conformal factor which must be positive to leave the signature of the metric unaltered. From this point forward
a tilde shall denote a quantity built out of the Einstein-frame metric tensor $\tilde g_{\m\n} $. Under this transformation the infinitesimal
line element and the determinant of the metric transform as
\be
d\tilde s^2 = \Omega^2 ds^2 \,, \qquad \sqrt{-\tilde g} = \Omega^4 \sqrt{-g} 
\,,
\ee
respectively. The Ricci scalar transforms as
\be
R = \Omega^2 \, \left( \tilde R + 6 \widetilde{\opensquare} (\ln{\Omega}) - 6 \tilde g^{\m\n} \frac{\widetilde \nabla_\m \Omega\widetilde \nabla_\n \Omega}{\Omega^2}\right)
\,.
\ee
Omitting the ordinary divergence $\sqrt{-\tilde g} \widetilde{\opensquare}  (\ln{\Omega}) $, the action in the Brans-Dicke frame transforms to
\be
\label{actcf}
S = \int d^4 \! x \, \sqrt{-\tilde g} \,\left(\frac{\Phi }{\Omega^2}\tilde R \,-\, 6 \frac{\Phi}{\Omega^4} \tilde g^{\m\n} \widetilde \nabla_\m \Omega \widetilde \nabla_\n \Omega + \frac{V(\Phi)}{\Omega^4} \right) 
\,,
\ee
where the potential in terms of $\Phi$ is given by (\ref{newp}) together with (\ref{vandf}):
\be
\label{vofbdp}
V(\Phi) = \frac{\m^3}{\g}\left(\frac{\mpl^2}{2} - \Phi \right)
\,.
\ee
Here we have re-introduced the parameter $\m$ from \eq{vandf} (previously set equal to $\g$) for completeness.

By choosing our conformal factor to be $\Omega^2 =2\Phi/\mpl^2$, and performing a field redefinition to define the Einstein frame field 
$\phi$, 
\be
\Phi = \frac{\mpl^2}{2} \exp{\left\{ {\sqrt{\frac{2}{3}}\frac{\phi}{\mpl}} \right\} }
\,,
\ee
the action (\ref{actcf}) becomes the Einstein frame action
\be
\label{acteinst}
S_{EF} = \int d^4 \! x \, \sqrt{-\tilde g} \,\left(\frac{\mpl^2}{2} \tilde R \,-\,  \half \tilde g^{\m\n} {\widetilde \nabla}_\m \phi \widetilde \nabla_\n \phi 
- \widetilde{V}(\phi) \right) 
\,,
\ee
where we have defined the potential
\be
 \widetilde{V}(\Phi) = - \left(\frac{\mpl^2}{2}\right)^2 \frac{V(\Phi)}{\Phi^2} 
 \,.
 \ee
The Einstein frame potential is given by
\be
\label{einstv}
\widetilde{V}(\phi) = \frac{\mpl^2}{2} \mathcal{R} \left(e^{\b \phi/\mpl} - 1\right) \, e^{-2 \b \phi/\mpl}
\,,
\ee
where $\mathcal R = \m^3/\g$ is the minimal curvature and $\b = \sqrt{2/3}$.

In this frame the field $\phi$ has a canonically normalized kinetic term making the interpretation of solutions more simple due
to our familiarity with minimally coupled scalar field to ordinary Einstein gravity. It is important to note, however, that so far we have only 
considered the vacuum theory. Even in the Einstein frame the Einstein field $\phi$ will couple non-minimally to matter and therefore,
when matter is present in significant amounts, the simple Einstein frame vacuum solutions will not be an accurate description of the
theory.

Under the conformal transformation (\ref{ctrans}), the cosmic time coordinate transforms as
\be\label{ttran}
d \tilde t^2 = e^{\b \phi/\mpl} \, dt^2
\,.
\ee
Taking the Einstein-frame Friedmann-Robertson-Walker flat metric
\be
d\tilde s^2 = - d \tilde t^2 + \tilde a^2(\tilde t) d\bf{x}^2
\,,
\ee
Leads to the familiar equations of motion
\be\label{sasha}
\tilde H^2 = \frac{1}{3 \mpl^2}\left( \half \phi'^2 + \tilde V(\phi) \right)
\,,
\ee
and 
\be
\phi'' + 3 \tilde H \phi' + \frac{d\tilde V}{d\phi} = 0
\,,
\ee
where prime denotes differentiation with respect to the Einstein frame cosmic time coordinate $\tilde t$ and $\tilde H \equiv \tilde a' / \tilde a$. Using the conformal transformation (\ref{ctrans}), we find the Hubble
parameters in the Einstein and Minimal Curvature frames are related by
\be\label{relate}
H = e^{\frac{\b \phi}{2\mpl}} \left( \tilde H - \frac{\b}{2\mpl} \phi' \right)
\,.
\ee

We plot the Einstein frame potential (\ref{einstv}) in \fig{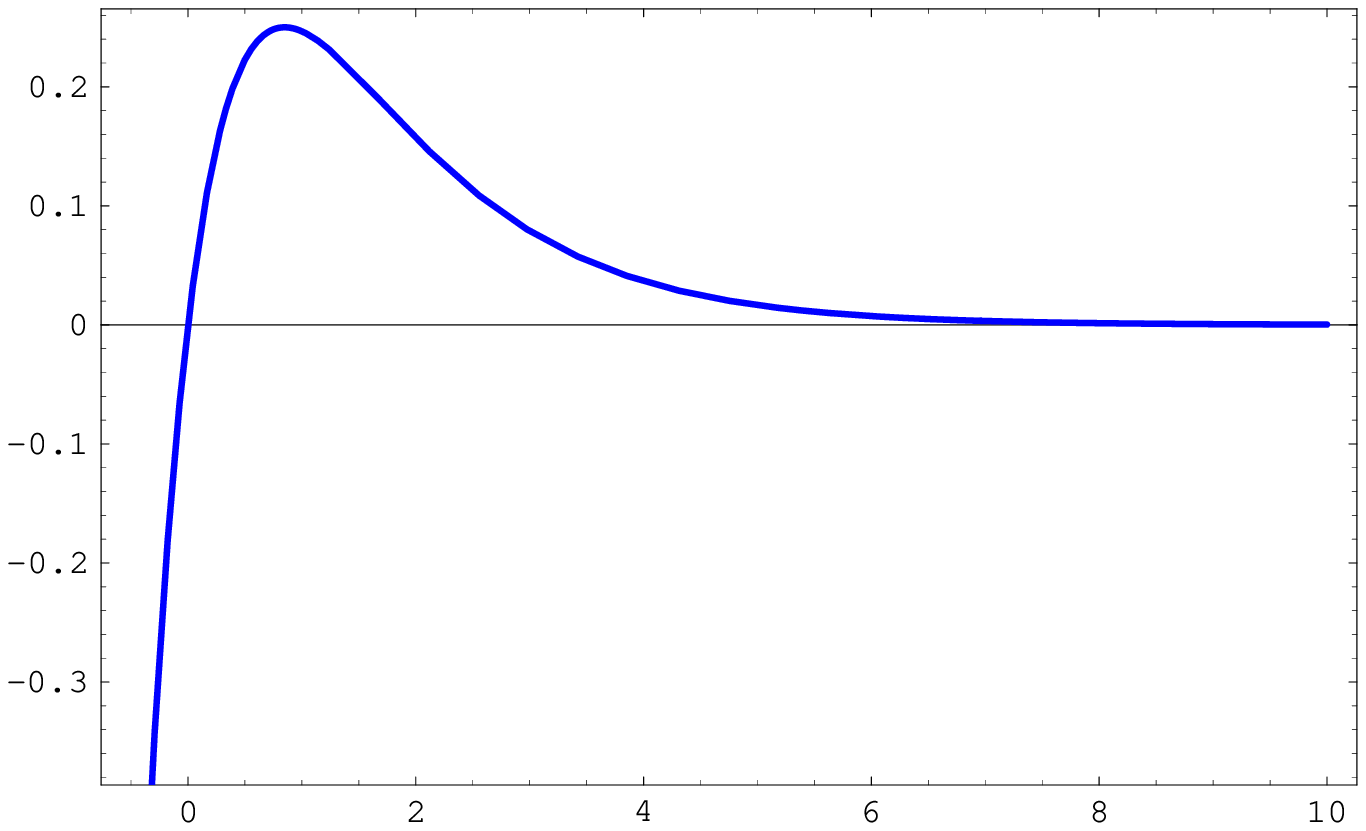}.
\begin{figure}[ht]
\begin{centering}
\includegraphics[width=5 in,clip]{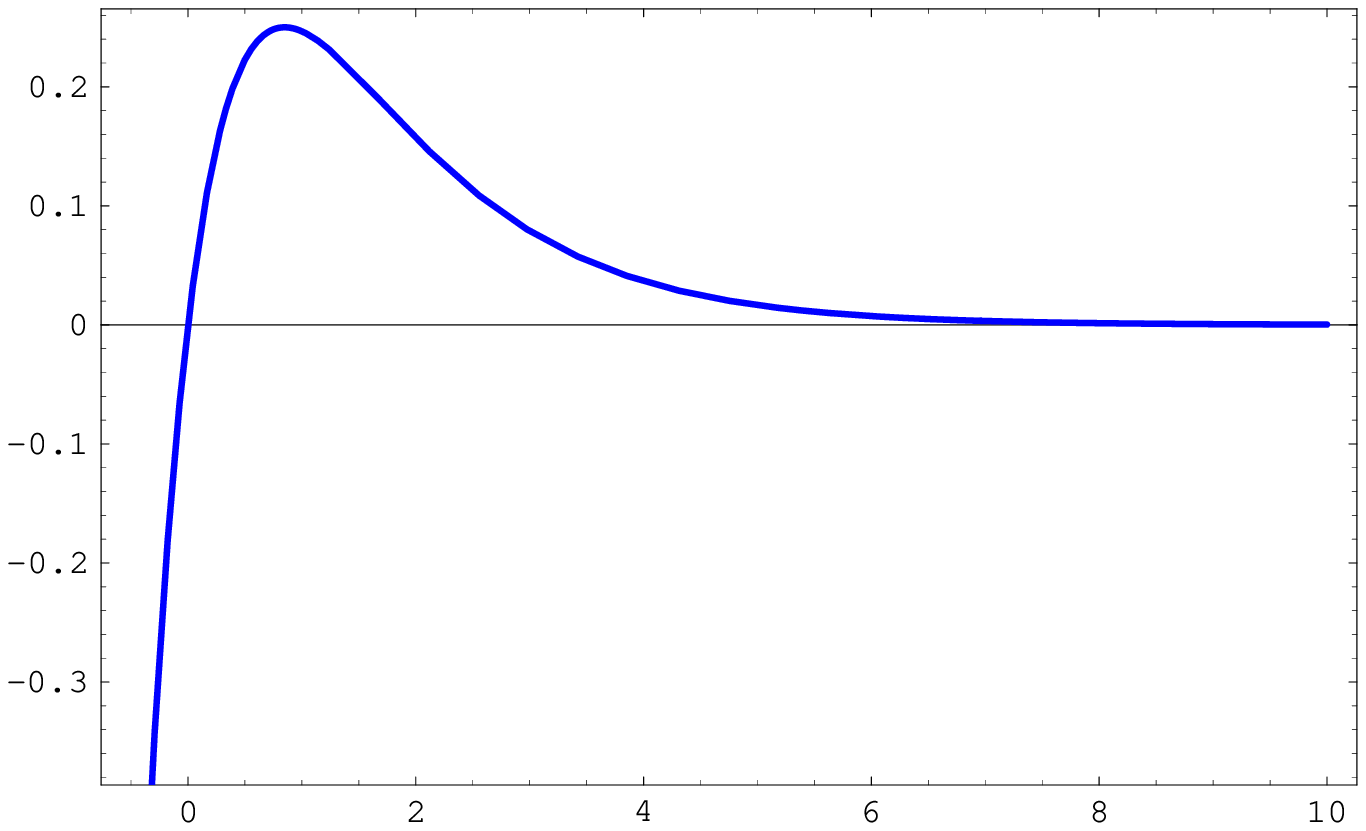}
\caption{The Einstein frame potential $\widetilde V$ as a function of the Einstein frame field $\phi$. In the plot we have set $16 \pi G_N= \g = \m =1$.}
\label{einstpot.eps}
\end{centering}
\end{figure}
From the plot it is clear that there are three regions of interest: 

i) There is an unstable de Sitter solution corresponding to the field
sitting on the top of the potential. 

ii) If the field starts initially to the left of the unstable de Sitter solution (and without significant positive velocity)
the system will quickly run off to large negative $\phi $. 

iii) If the field falls to the right of the de Sitter solution it will asymptotically roll to large positive values.

Let us examine the first possibility in greater detail. The unstable de Sitter solution with constant $\tilde H = \tilde H_{dS}$ sits at the value at the maximum of the
potential (\ref{einstv}). The maximum is at constant value 
\be\label{dspt}
\phi = \phi_{dS} = \b^{-1} \mpl \ln{2}
\,,
\ee
From \eq{sasha} we find the corresponding $\tilde H$
\be
\tilde H_{dS} = \half \sqrt{\frac{\mathcal R}{6}}
\,.
\ee
From \eq{relate} we see that the corresponding value in the minimal curvature frame is
\be
H(\tilde H_{dS}, \, \phi_{dS}) = \sqrt{\frac{\mathcal{R}}{12}}
\,,
\ee
which is the exact de Sitter solution we found in our analysis of the vacuum of the minimal curvature theory (\ref{julia}).

The Minimal Curvature frame field is related to the Einstein frame field via
\be\label{vpofphi}
\vp(\phi) = \frac{\mpl^2}{2\g} \left(1 - e^{\b \phi/\mpl} \right)
\,,
\ee
and consequently,
\be
\phi(\vp) = \frac{\mpl}{\beta} \ln{\left( 1 - \frac{2\g}{\mpl^2}\vp \right)}
\,.
\ee
Using the relation (\ref{vpofphi}), we find the value of $\vp$ at the de Sitter point (\ref{dspt}):
\be
\vp(\phi_{dS}) = - \frac{\mpl^2}{2\g}
\,.
\ee
Hence we conclude that the unstable de Sitter solution in the Einstein frame is mapped to one of the unstable saddle critical points in the vacuum of the
Minimal Curvature frame (see \eq{eqpts}). 

We now examine case iii): when $\phi$ rolls to large positive values the Einstein frame potential may be approximated by
\be
\label{vprox}
\widetilde{V}(\phi) \simeq \frac{\mpl^2 \mathcal{R}  }{2} e^{-\b \phi/\mpl}
\,,
\ee
leading to the exact solutions
\bea
\frac{\phi(\tilde t)}{\mpl} = \sqrt{6} \ln{\left(\sqrt{\frac{\widetilde V_0}{24}} \frac{\tilde t}{\mpl} \right)} \\
\tilde H(\tilde t) = \frac{3}{\tilde t}
\,,
\eea
where $\widetilde{V_0} = \mpl^2 \mathcal R/2$, corresponding to power-law acceleration in the Einstein frame with
\be\label{ltpl}
\tilde a(\tilde t) = \tilde a_0 \tilde t^3
\,.
\ee
To find the corresponding behavior in the Minimimal Curvature frame we use the conformal transformation (\ref{ctrans}) along with
the transformation for the cosmic time coordinate~(\ref{ttran}). We find the late-time power law attractor in the Einstein frame (\ref{ltpl})
is mapped to the asymptotic de Sitter attractor in the MC frame at the minimal curvature $\mathcal R$ (depicted by the dashed
blue line in \fig{vacphase.eps}),
\be
a(t) = a_0 e^{\sqrt{\frac{{\mathcal R}}{12}} t}
\,.
\ee

\end{document}